# Terahertz sensing of highly absorptive water-methanol mixtures with Fano resonances in metamaterials

Min Chen,[1,2] Leena Singh,[2] Ningning Xu,[2] Ranjan Singh,[3,4] Lijuan Xie,[1,a)] and Weili Zhang[2,a)]

[1]*College of Biosystems Engineering and Food Science, Zhejiang University, 866 Yuhangtang Road, Hangzhou 310058, PR China*

[2]*School of Electrical and Computer Engineering, Oklahoma State University, Stillwater, Oklahoma 74078, USA*

[3]*Division of Physics and Applied Physics, School of Physical and Mathematical Sciences, Nanyang Technological University, 21 Nanyang Link, Singapore 637371, Singapore*

[4]*Center for Disruptive Photonic Technologies, The Photonics Institute, 50 Nanyang Avenue, Nanyang Technological University, Singapore 639798, Singapore*

Ultrasensitive terahertz sensing of highly absorptive aqueous solutions remains challenging due to strong absorption of water in the terahertz regime. Here, we experimentally demonstrate a cost-effective metamaterial based sensor integrated with terahertz time-domain spectroscopy (THz-TDS) for highly absorptive water-methanol mixture sensing. This metamaterial has simple asymmetric wire structures that supports a fundamental Fano resonance and higher order dipolar resonance in the terahertz regime. Both the resonance modes has strong intensity in the transmission spectra which we exploit for detection of highly absorptive water-methanol mixtures. The experimentally measured sensitivities of the Fano and the dipole resonances for water-methanol mixtures are found to be 169, and 308 GHz/RIU, respectively. Such an ultrasensitive sensing provides a route for readily available metamaterial-assisted terahertz spectroscopy for ultrasensitive sensing of highly absorptive chemical and biochemical materials.

Metamaterials are artificially engineered materials with periodically arranged, sub-wavelength structures and exhibit unique electromagnetic properties that are unavailable among the naturally existing materials.[1-7] Due to the strong localization and enhancement of the electromagnetic fields leading to fertile hot-spots, metamaterials exhibit strong optical response towards the presence of an analyte. Such a response leads to material sensing with metamaterials in different spectral range, extending from microwaves to optics.[8-15]

Recently, metamaterial-assisted sensing platform integrated with terahertz spectroscopy has attracted a lot of interests and is being increasingly implemented for chemical and biological sensing.[11,12,16-30] The key advantages of this sensing platform include high sensitivity, real-time, and label-free detection abilities. However, at terahertz frequencies, owing to the strong absorption of polar liquids, such as water, most of these studies have typically been limited to dry or partially hydrated specimens. Very few reports on direct liquid sensing based on terahertz metamaterials have been presented.[11,17,20,22,23,27] Since most of the functionalities of chemical and biological materials are expressed in water, the key solvent of most biological substances, therefore, it is important to realize readily available and cost-effective sensing platforms for water-based real biological systems.

In this article, we provide a novel, cost-effective metamaterial sensor with a simple metallic structure but strong and sensitive multi-modal resonance based terahertz sensing for highly absorptive water-methanol mixtures. The metamaterial sensor consists of asymmetric dual wire arrays which offers the ease of fabrication and exhibits multi-modal resonances including a Fano resonance dip, a dipole resonance dip, and a Fano transmittance peak between the two resonance frequencies due to the structural symmetry breaking. Predominantly, both resonances have strong intensity and are able to sense highly absorptive solutions. In addition, we use free standing, transparent and thin Mylar substrate with a total thickness of 22 μm. The sensitivity is enhanced due to the fact, that the thin substrate, with lower dielectric constant, has lower influence on the capacitance of the resonators.[31-35] We successfully demonstrate that the proposed metamaterial sensor combined with terahertz time-domain spectroscopy (THz-TDS) could be effectively utilized in terahertz sensing of water-methanol mixtures with the capability to identify solvent type and determine the corresponding concentrations. This work would motivate metamaterial-assisted terahertz sensing for chemical and biological substances in highly absorptive aqueous systems.

Figure 1(a) shows the microscopic image of the proposed asymmetric dual wire resonator (ADWR) with the inset showing the unit cell schematic. The unit cell structure of ADWR is composed of two aluminum (Al) wires placed in parallel on the Mylar substrate. The detailed structural parameters are as follows: $P_x$ = 80 μm, $P_y$ = 126 μm, $L_1$ = 60 μm, $L_2$ = 106 μm, $d$ = 20 μm, and $\omega$ = 6 μm. The thickness of the Al and Mylar films is 200 nm and 22 μm, respectively.

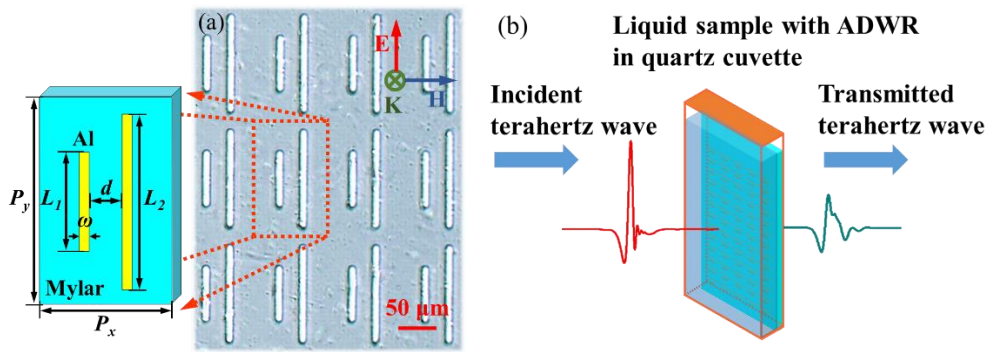

FIG. 1. (a) Microscopic image of the ADWR with the inset showing the schematic unit cell structure where $P_x$ = 80 μm, $P_y$ = 126 μm, $L_1$ = 60 μm, $L_2$ = 106 μm, $d$ = 20 μm, and $\omega$ = 6 μm. (b) Schematic diagram of liquid sample measurement by ADWR combined with THz-TDS.

Numerical simulation for ADWR to obtain transmission parameters was carried out using CST Microwave Studio frequency domain solver with tetrahedral mesh.[36] Mylar substrate was modeled as a lossless dielectric with dielectric permittivity $\varepsilon$ = 2.89 and the DC conductivity $\sigma$ of Al metal was set to be 3.56 × 10$^7$ S/m. The ADWR was fabricated as follow: dual wire structures were first patterned on Mylar substrate by conventional photolithography using positive photoresist. A 200-nm-thick Al film was then thermally deposited and followed by lift-off process to form the wire structures.

A traditional photoconductive switch-based THz-TDS system was used to measure the transmitted terahertz signal through the samples and references.[20,37] Figure 1(b) shows the schematic diagram of liquid sample measurement using ADWR-assisted THz-TDS. The ADWR was placed in a quartz cuvette composed of two parallel, 60-μm-thick spaced, and 1-mm-thick windows with the side of the Mylar substrate closely attaching one quartz window. The liquid sample was then injected into the cuvette. The cuvette was sealed and placed midway between the transmitter and receiver in the focused beam of THz-TDS systems to obtain the transmitted spectra. The polarization of the incident terahertz field was parallel to the wires. An identical empty quartz cuvette was used as the reference. The amplitude transmission $t(\omega)$ was obtained by using $t(\omega) = |\tilde{E}_s(\omega) / \tilde{E}_r(\omega)|$, where $\tilde{E}_s(\omega)$ and $\tilde{E}_r(\omega)$ are fast Fourier transformed (FFT) transmitted electric field of the sample and the reference pulses, respectively, $\omega$ is the angular frequency. The average of three times' repeated transmission measurements was used for further analysis.

Double-distilled water (home-made) and methanol with a purity higher than 99.9% (Pharmco-AAper) were used in this work without further purification. The water-methanol mixtures were prepared with 0%, 30%, 50%, 70%, 100% (v/v %) water volume percentage.

Figure 2(a) presents the simulated amplitude transmission of the ADWR and the two individual wire structures with $L_1$ = 60 μm, $L_2$ = 0 μm, and $L_1$ = 0 μm, $L_2$ = 106 μm. The two individual wire structures resonated at two different frequencies due to their unequal wire lengths. The short wire structure resonated at high frequency, $f$ = 1.695 THz, while the longer wire structure resonated at low frequency, $f$ = 1.068 THz. However, once the two wires were placed together, the resonance in the coupled ADWR exhibited multi-modal resonances including an asymmetric resonance dip at 1.038 THz, a transmittance peak at 1.269 THz, and a symmetric resonance dip at 1.734 THz in the transmission spectra. It should be noted that the two resonance dips of the coupled ADWR happened to be located near the two resonance dips of the individual wire arrays. The insets of Figure 2(a) present the surface current distributions at the two resonance dips of the coupled ADWR. We observed Fano-like antiparallel currents at 1.038 THz, and a dipole-like parallel currents at 1.734 THz.

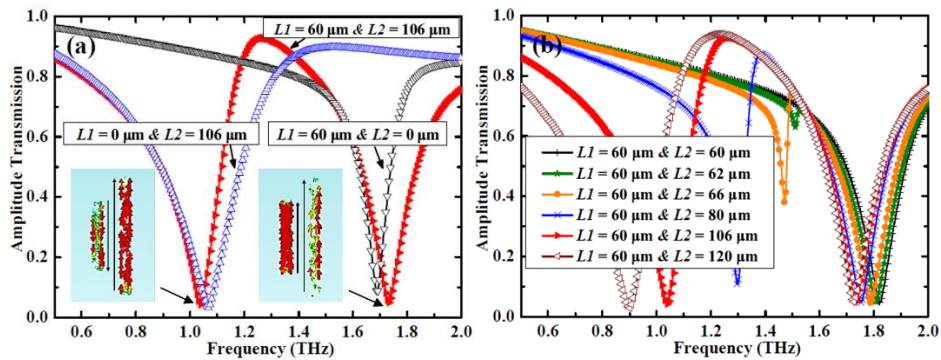

FIG. 2. Simulated amplitude transmission spectra of (a) the ADWR with $L_1$ = 60 μm, $L_2$ = 106 μm and the two individual wire structures with the insets depicting the surface current distributions of the ADWR at 1.038 and 1.734 THz, and (b) six DWRs with different asymmetry.

In order to understand the mechanism of these multi-modal resonances especially the Fano resonance, we investigated the transmission responses of six dual wire resonators (DWRs) with different asymmetry which are defined as the length differences between the two wires in the unit cell. Figure 2(b) shows the simulated amplitude transmission spectra of the six DWRs, including one perfect symmetric dual wire resonator (SDWR) with $L_1 = L_2 = 60$ μm and five ADWRs with $L_2$ increasing from 60 to 120 μm but constant $L_1$ = 60 μm. The other parameters of these six DWRs are constant with $P_x$ = 80 μm, $P_y$ = 126 μm, $d$ = 20 μm, and $\omega$ = 6 μm. For the SDWR, there was only a single resonance at 1.824 THz observable in the transmission spectrum. However, when the two wires had different lengths, that is, the structural asymmetry was introduced, multi-modal resonances could be observed. In addition, a gradual red shift was observed in the frequency positions of both the resonances, and the Fano resonances became significantly broader but stronger with the increase of the asymmetry.

These phenomena could be contributed to the destructive interference between the two wires caused by symmetry breaking.[38-41] In the symmetric system, there was no destructive interference between the two wires, and hence only one broad dipole resonance in the transmission spectra is seen. However, once the structural symmetry is broken in the unit cell with two unequal lengths of wires, the destructive coupling at the lower frequency leads to antiparallel surface current distribution that gives rise to the sharp Fano resonance. The surface current distribution at the higher frequency resonance is parallel in nature which gives rise to the dipolar resonance mode as shown in the inset of Figure 2 (a). When the length difference between the two wires increased, the $Q$ factor of the Fano resonance dip declined with an enhancement of the resonance intensity was enhanced as shown in Figure 2(b).

We further investigated the sensitivity in terms of the change in resonance frequency per refractive index unit (RIU) of the Fano resonance dip of the five ADWRs shown in Figure 2 (b) using rigorous simulations. The resonance frequency of the Fano resonance dip for 4 μm analyte thickness while varying the refractive index from $n$ = 1.0 to $n$ = 1.6 in incremental steps of 0.2 was calculated by CST. Table 1 lists the simulated Fano resonance sensitivity and intensity of the five ADWRs. The Fano resonance intensity here was defined as the power transmission (the square of the amplitude transmission) difference between the Fano resonance dip and the Fano transmittance peak.

TABLE I. Summary of Fano resonance sensitivity and intensity of the five ADWRs with varying $L_2$ shown in Figure 2(b).

|  | $L_2$ = 62 μm | $L_2$ = 66 μm | $L_2$ = 80 μm | $L_2$ = 106 μm | $L_2$ = 120 μm |
|---|---|---|---|---|---|
| Sensitivity (GHz/RIU) | 268 | 266 | 217 | 176 | 161 |
| Intensity | 0.075 | 0.414 | 0.744 | 0.861 | 0.877 |

When the length difference between the two wires was increased, the Fano resonance sensitivity gradually decreased from 268 to 161 GHz/RIU, while the intensity of Fano resonance increased from 0.075 to 0.877, as shown in Table I. It should be noted that strong resonance intensity is extremely necessary for this resonance to be measured when dealing with strong absorptive liquid samples, especially when the measurement is performed with low resolution and low signal-to-noise ratio systems. Though the Fano resonance sensitivities of ADWRs with $L_2$ below 106 μm were higher, their intensities were below 0.8. Therefore, we selected the ADWR with $L_1 = 60$ μm, $L_2 = 106$ μm as the terahertz metamaterial for further experimental study of water-methanol mixtures. It should be noticed that the dipole resonance intensity of this selected ADWR was also strong. In addition, the simulated Fano resonance sensitivity of this ADWR (176 GHz/RIU) was almost 5 times that of the terahertz asymmetric split ring (36.7 GHz/RIU) shown in previous report.[26]

We firstly measured the terahertz transmission results of water-methanol mixtures in quartz cuvette without ADWR and extracted the refractive indices of different water contents for subsequent measurements with ADWR. Figure 3(a) depicts the measured frequency dependent amplitude transmission results of the water-methanol mixtures with the water contents changing from 0% (pure methanol) to 100% (pure water) at 0.5-2 THz of interest. The measured amplitude transmission of different water-methanol mixtures decreased with the increase of the water content due to the stronger absorption of water than methanol. The refractive indices $n(\omega)$ of different water-methanol mixtures was derived from the transmission results by using $n(\omega) = |\Phi_s - \Phi_r| \times c / (\omega \times d) + n_r$,[42,43] where $\Phi_s$ and $\Phi_r$ are the phases of the sample and reference, respectively, $c$ is the speed of light, $\omega$ is the angular frequency, $d$ is the thickness of the liquid sample, and $n_r$ is the refractive index of the reference. Figure 3(b) presents the frequency dependent refractive indices $n(\omega)$ of different water-methanol mixtures. The refractive indices of the water-methanol mixtures increased with the increase of water content due to the higher refractive index of water. The refractive indices of pure methanol and pure water as shown in Figure 3(b) agreed with the previous works.[44,45] The measured average refractive indices of water-methanol mixtures with 0%, 30%, 50%, 70%, and 100% water contents were 1.86, 2.17, 2.29, 2.44, and 2.59, respectively.

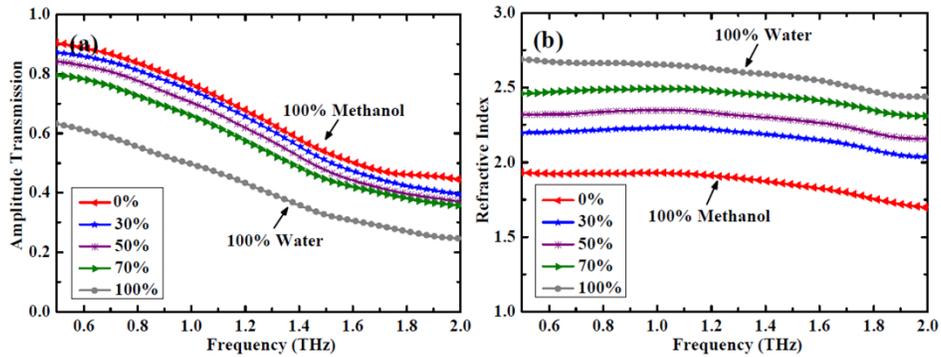

FIG. 3. (a) Measured amplitude transmission spectra and (b) refractive indices of water-methanol mixtures with different water contents from 0% to 100%.

Subsequently, we measured the water-methanol mixtures with ADWR. Figure 4(a) shows the measured amplitude transmission spectrum of the bare ADWR metamaterial. The simulated amplitude transmission of it was also shown for comparison. We can observe that the experimental amplitude transmission of the bare ADWR is in good agreement with the simulated amplitude transmission. In order to enlarge spectra variations with different water contents, we normalized the amplitude transmission of water-methanol mixtures obtained by the ADWR-assisted THz-TDS with the terahertz transmission of 100% methanol without ADWR metamaterial shown in Figure 3(a). Figure 4(b) presents the normalized amplitude transmission spectra of water-methanol mixtures with ADWR. A gradual red shift was observed at both resonances (Fano resonance dip, Fano transmittance peak, and dipole resonance dip) with the increase of water contents or refractive indices.

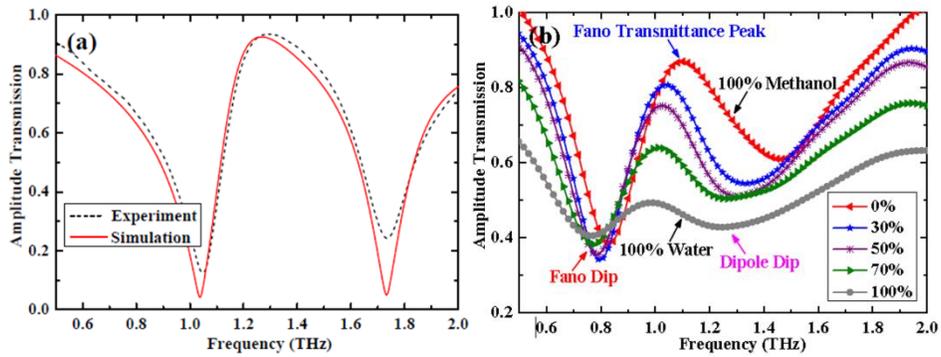

FIG. 4. (a) Experimental and simulated amplitude transmission spectra through the bare ADWR metamaterial. (b) The normalized measured amplitude transmission spectra of water-methanol mixtures obtained by ADWR-assisted THz-TDS.

We further plotted the resonance frequency shifts of both resonances with the changes in mean refractive indices of the water-methanol mixtures in Figure 5 (a). Sensitivities of the Fano resonance dip, the Fano transmittance peak, and the dipole resonance dip turned out to be 97, 169, and 308 GHz/RIU, respectively. More importantly, the amplitude of the Fano transmittance peak fitted exponentially with the changes of the refractive indices or the concentrations of water in the mixtures, and their resonance frequencies were linear fitted with the changes of refractive indices, as shown in Figure 5(b). We could realize qualitative and quantitative sensing of methanol and water.

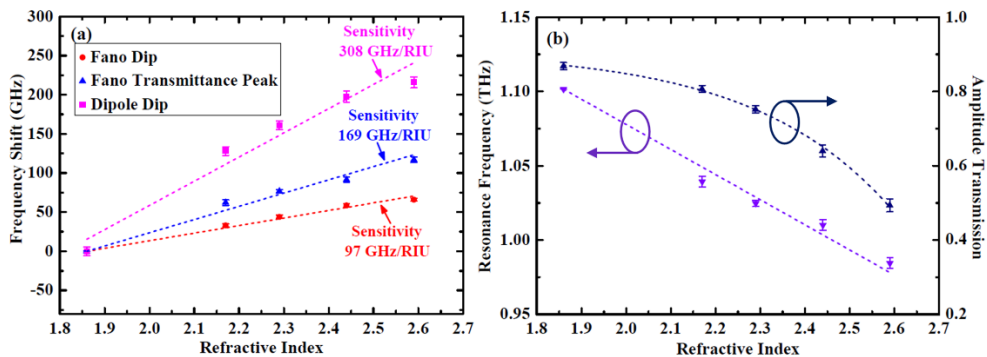

FIG. 5. Measured frequency shifts of (a) Fano resonance dip, Fano transmittance peak, and dipole resonance dip with the changes of refractive indices of different water contents. (b) Resonance frequencies and transmission values of the Fano transmittance peak with the changes of refractive indices of different water contents.

In conclusion, by breaking the symmetry of dual wire structures on Mylar substrate via their unequal lengths, we obtained a readily available metamaterial with multi-modal resonances in terahertz regime including a fundamental Fano resonance dip, a dipole resonance dip, and a Fano transmittance peak between them. By further optimizing the length difference between the two wires, a practicable terahertz metamaterial sensor with strong multi-modal resonances and high sensitivities was achieved for efficient sensing of strongly absorptive water-methanol mixtures. The experimental results of water-methanol mixtures sensing show that the designed metamaterial can achieve frequency sensitivities of 97, 169, and 308 GHz/RIU for the Fano resonance dip, the Fano transmittance peak, and the dipole resonance dip, respectively. This metamaterial presents a promising avenue for cost-effective, label-free, real-time, qualitative and quantitative terahertz sensing of chemical and biological substances in water, methanol, and other kinds of highly absorptive aqueous systems.

This work was partially supported by the National Natural Science Foundation of China (Grant No.31471410). This work was performed in the Ultrafast THz Optoelectronic Laboratory at Oklahoma State University in Stillwater, Oklahoma, USA when Min Chen was visiting there on a joint Ph.D. degree program under the sponsorship of China's Scholarship Council (CSC).